\documentclass[aps,pra,superscriptaddress,twocolumn]{revtex4-1}
\usepackage{braket}
\usepackage{amsmath}
\usepackage{natbib}
\usepackage{graphicx}
\newcommand*{\cra}{^\dag}
\begin{document}
\title{Linear Optical Deterministic and Reconfigurable SWAP Gate}
\date{\today}
\author{Alkım B. Bozkurt}
\affiliation{Electrical and Electronics Engineering Department,
Middle East Technical University, 06800 Ankara, Turkey}
\author{Serdar Kocaman}
\email{skocaman@metu.edu.tr}
\affiliation{Electrical and Electronics Engineering Department,
Middle East Technical University, 06800 Ankara, Turkey}
\begin{abstract}
We propose a deterministic SWAP gate for spatially encoded qubits. The gate is constructed from waveguide crossings, Mach Zender Interferometers and phase shifters providing the gate reconfigurability. Through manipulating the phase of the phase shifters, we can apply either the SWAP or identity gates. As an essential element of nearest neighbor qubit networks, the SWAP gate has a simple structure and presents minimal overhead, making it viable for universal quantum computation.
\end{abstract}
\maketitle
\section{Introduction}

Manipulating quantum wavefunctions for information processing, quantum information science introduces significant enhancements to computation, communication and metrology \cite{Nielsen2002,Crypto1,Giovannetti2011}. Execution of algorithms offering considerable speed up compared to their classical counterparts and achieving universal quantum computation is a fundamental goal of quantum information science.
Single photon qubits are appealing candidates for the implementation of quantum information technologies due to their low decoherence and flying nature \cite{o2009photonic}. On-chip quantum circuits for spatially encoded photonic qubits are particularly promising due to their stability, scalability and robustness \cite{Politi2009,Matthews2009,Silverstone2016,Carolan2015}. Silicon photonics, with its CMOS compatibility, rapid plasma dispersion effect modulators which provide reconfigurability and the possible integration of sources, photodetectors and electronics on-chip, is a powerful medium for the production of devices performing quantum computation tasks \cite{Thompson2012}.

General quantum computation algorithms assume that it is possible to interact any two qubits regardless of their locations \cite{Nielsen2002}. However, leading physical implementations of systems capable of processing quantum information can only have nearest neighbor interactions where circuits are mostly composed of single and two qubit operations \cite{Ladd2010}. Although it is theoretically possible to interact two arbitrary qubits by applying multiple SWAP gates which place them in neighboring positions, the associated overhead and costs are high. The physical restriction of nearest neighbor qubit interactions has caused the emergence of new algorithms optimized for nearest neighbor networks \cite{Fowler2004,Takahashi:2007:QFT:2011725.2011732,Fowler2004a}. It is necessary to design circuits with the least possible number of SWAP gates and to physically implement compact SWAP gates for scalability.

The SWAP gate is given by the unitary transformation 
\begin{equation}
\label{eqn:SWAP}
	\text{SWAP} =
	\begin{pmatrix}
		1 & 0 & 0 & 0 \\
		0 & 0 & 1 & 0 \\
		0 & 1 & 0 & 0 \\
		0 & 0 & 0 & 1 
	\end{pmatrix}
\end{equation}
Acting on an arbitrary product state
\begin{equation}  \label{eqn:2}
	\psi _1 \otimes \psi _2 \xrightarrow{\text{SWAP}} \psi _2 \otimes \psi _1
	\end {equation}

\begin{figure}
\includegraphics[width=0.85\columnwidth]{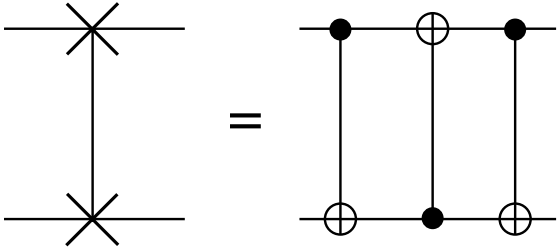}
\centering
\caption{Circuit representation of the SWAP gate showing its equivalence to three consequtive CNOT gates.}
\label{gatecircuit} 

\end{figure}

The common scheme to apply the SWAP gate consists of three successive CNOT gates as seen in Fig. \ref{gatecircuit} \cite{Nielsen2002}. Whereas this approach is plausible for most physical implementations \cite{Schmidt-Kaler2003,Zajac2018,Plantenberg2007}, it represents a significant overhead for linear optical quantum computation, where entangling operations such as CNOT can only be realized probabilistically \cite{Knill2001}. With on-chip circuits, there have been demonstrations of an unscalable, unheralded CNOT gate with success probability of $ 1/9 $, requiring no ancilla qubits  \cite{Ralph2002,Politi2008} and a heralded CNOT gate with success probability of $ 1/16 $, requiring two photon detectors and an auxiliary photon, which is scalable due to the heralded nature of the gate \cite{Knill2001,Carolan2015}. Despite the progress, implementation of a SWAP gate through successive CNOT gates is inherently unscalable since it can only be attained with an extremely low probability ( $ 1/4096 $ for the heralded gate ) and a significant resource overhead.
As clearly seen in Eq.\eqref{eqn:2} the SWAP gate is not an entangling gate; hence it may be possible to construct it in a deterministic manner, thereby surpassing the restrictions associated with entangling operations.

It is possible to consider SWAP as a gate that switches $ \ket{0}_1 $ with $ \ket{0}_2 $ and $ \ket{1}_1 $ with $ \ket{1}_2 $. Although operating on the logical basis individually is highly challenging for many implementations of quantum computation \cite{Ladd2010}, the physically seperate nature of the logical basis on spatially encoded qubit networks easily permit such an approach. We propose a deterministic and reconfigurable SWAP gate by exploiting the distinction of the waveguides, which carry photons corresponding to different logical basis.

\section{Design} \label{design}
\begin{figure*}[!htb]
\includegraphics[width=.95\linewidth]{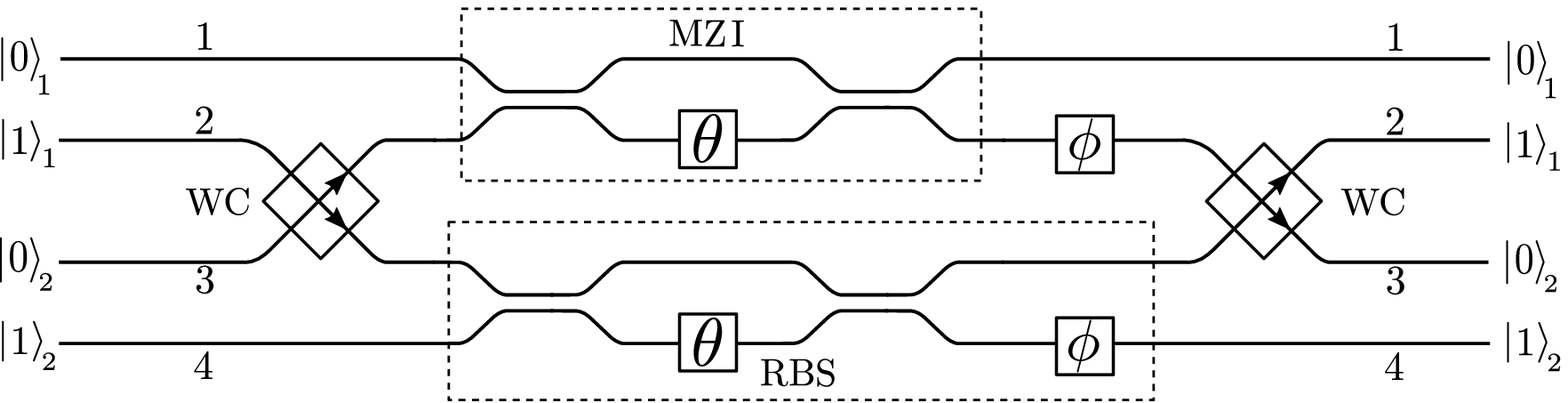}
\centering
\caption{ Schematic of the proposed SWAP gate. Waveguide crossings  are labeled WC, Mach Zender Interferometers (MZI) are composed of $ \eta = 0.5 $ directional couplers (or MMI couplers) and internal phase shifter  $\theta $. The reconfigurable beam splitter (RBS) is composed of an MZI and an external phase shifter $ \phi $.}
\label{gatefig}
\end{figure*}

Since on-chip photonic networks consist of nearest neighbor qubits, a method must be devised to bring the corresponding logical basis of two qubits next to one another. This can be done through waveguide crossings (denoted as WC) which cross $ \ket{1}_1 $ and $ \ket{0}_2 $, after which we obtain the desired configuration, where the logical basis to be swapped are neighboring (see Fig. \ref{gatefig}). Reconfigurability is essential for universal quantum computation; hence we wish to execute the swapping in a reconfigurable manner, where the switching is performed to obtain the SWAP gate and the identity operation is performed to obtain the overall 4x4 identity gate. This reconfigurability can be obtained by using a reconfigurable beam splitter (RBS),  which consists of a Mach Zender Interforemeter with an internal phase shifter $ \theta $, followed by an external phase shifter $ \phi $. The internal and external phase shifters are located at the same rail (Fig. \ref{gatefig}). The resultant structure performs the unitary transformation 

\begin{equation}
\label{eqn:3}
\text{U}_{\text{RBS}}(\theta,\phi) =
\begin{pmatrix}
sin({\theta /2}) & cos({\theta /2}) \\
e^{i\phi}cos({\theta /2}) & -e^{i\phi}sin({\theta /2})
\end{pmatrix}
\end{equation}
Substituting $ \theta=0 $ and $ \phi=0$ in Eq. \eqref{eqn:3} we obtain the unitary transformation corresponding to the Pauli X gate, which will be swapping the logical basis. When $ \theta=\pi $ and $ \phi=\pi$, we obtain the identity operation, which will provide that no swapping is performed. After the RBS transformation, we use waveguide crossings to bring the qubits to the spatially encoded arrangement, back into the logical state-space and finalize the gate.

We now demonstrate the operation of our gate starting with an arbitrary state $ \alpha \ket {00} + \beta \ket {01} + \gamma \ket {10} + \delta \ket {11} $ where $|\alpha| ^2 +|\beta|^2+|\gamma|^2+|\delta|^2=1 $ for normalization and setting $ \theta=0 $ and $ \phi=0$ . We may represent this state through creation-annihilation operators, where the subscripts denote the rails as labeled in Fig. \ref{gatefig} and $ \ket{0}$ is the vacuum state  as:

\begin{align}
& \left(  \alpha \,  a_1\cra a_3\cra + \beta \, a_1\cra a_4\cra + \gamma \,a_2\cra a_3\cra + \delta\, a_2\cra a_4\cra \right)\ket{0} \\
\xrightarrow{\text{WC}} & \left(  \alpha\,  a_1\cra a_2\cra + \beta\, a_1\cra a_4\cra + \gamma\, a_2\cra a_3\cra + \delta\, a_3\cra a_4\cra \right)\ket{0} \label{WC1}\\
\xrightarrow{\text{RBS}} & \left(  \alpha\,  a_1\cra a_2\cra + \beta\, a_2\cra a_3\cra + \gamma\, a_1\cra a_4\cra + \delta\, a_3\cra a_4\cra \right)\ket{0} \\
\xrightarrow{\text{WC}} & \left(  \alpha\,  a_1\cra a_3\cra + \beta\, a_2\cra a_3\cra + \gamma\, a_1\cra a_4\cra + \delta\, a_2\cra a_4\cra \right)\ket{0} \label{WC2}
\end{align}
We end up with the final state  $  \alpha \ket {00} + \beta \ket {10} + \gamma \ket {01} + \delta \ket {11} $, in Eq. (\ref{WC2}), we can observe that the operation of our gate corresponds to the unitary transformation of the SWAP gate given in Eq. \eqref{eqn:SWAP}. Thus we have successfully implemented the SWAP gate for $ \theta=0 $ and $ \phi=0$ .

The case when $ \theta=\pi $ and $ \phi=\pi$ is straightforward to verify. Since the RBS performs the identity operation, only the action of the waveguide crossings are significant. It can be seen from Eqs. (\ref{WC1},\ref{WC2}) that the action of WC is to swap $ a_2\cra $ and $ a_3\cra $, hence it is its own inverse. Applying WC twice, we get the overall identity operation and illustrate that our gate can apply the SWAP or identity gates in a reconfigurable and deterministic manner.

It is clearly seen from Fig. \ref{gatefig} that the proposed gate is asymmetric. We apply the WC operation to two photons, whereas the other two photons remain unaffected. Phase matching is necessary for the correct operation of our gate, otherwise the fidelity is likely to decrease significantly; hence consideration must be given to prevent phase matching  complications due to the asymmetry. The effective refractive index of the fundamental mode of the waveguide crossing must be taken into account and the path of the unaffected photons must be adequately modified to attain phase matching. Furthermore, temporal matching is required to preserve quantum interference, whereas a time delay causes distinguishability between photons  \cite{PhysRevLett.59.2044,Politi2008}. Physical design of the gate for avoiding phase and temporal mismatch may be significantly simplified by connecting the unaffected photons to waveguide crossings coupled to unoccupied waveguides. This simplification comes with a trade-off of resource overhead and an increase in the losses due to the WC. However, the multi-mode nature of most waveguide crossings can introduce temporal mismatch due to the varying refractive indices of the multiple modes, which has been associated with decreased visibilities of structures composed of MMI couplers \cite{Peruzzo2011}. It is possible to prevent the issues associated with MMI based waveguide crossings by replacing them with an MZI with an internal phase of $ \theta=0 $ (Eq. \ref{eqn:3}), i.e. an MZI without  an internal phase shifter. Although the MZI is less compact, its operation is likely to be more reliable thanks to the lack of multimode effects; furthermore it can significantly simplify the design required for phase and temporal matching.

\section{Selective Measurements}

\begin{figure}[!b]
	\includegraphics[width=0.75\columnwidth]{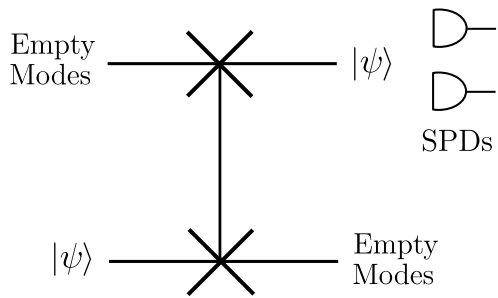}
	\centering
	\caption{ Circuit representation of the proposed selective measurement scheme where a measurement takes place. $\ket{\psi}$ is the qubit of interest and SPDs are single photon detectors.}
	\label{mescircuit}
\end{figure}

Measurements in the logical basis are performed by single photon detectors coupled to waveguides at the end of the circuit for on-chip devices. A shortcoming of this approach is that the measurement structure is application specific if feedback is to be implemented and it lacks reconfigurability. It is desirable to measure any qubit at the logical basis at any point in the circuit for universal quantum computation.

It is possible to utilize the proposed SWAP gate to make measurements selectively. Selective measurement means that one can measure a qubit or alternatively apply the identity operation without performing a measurement. The scheme requires empty modes (unoccupied waveguides) coupled to single photon detectors. When a measurement is desired, the empty modes are swapped with the qubit to be measured, using the proposed SWAP gate, as seen in Fig. \ref{mescircuit}. After swapping, the qubit of interest is measured by the  detectors. Rather than the SWAP operation, if we apply the identity operation via the proposed reconfigurable gate, we observe that no measurement takes place and the qubit of interest remains on its path undisturbed. Therefore, by manipulating the phase shifters of our gate it is possible to couple the qubit to the detectors and perform a measurement or instead apply the identity operation if a measurement isn't desired.

With respect to the the discussion in Sec. \ref{design}, we may assume that the main error in the gate is introduced through a phase mismatch. When a measurement is to be performed on an arbitrary qubit $ \ket{\psi}= \alpha \ket{0} + \beta \ket{1} $ with $|\alpha|^2+|\beta|^2=1$ satisfying normalization, we obtain $\ket{\psi}=\alpha \, e^{i\eta} \ket{0} + \beta \ket{1}$ at the output port, where the qubit is coupled to the photodetectors, $ e^{i\eta}$ representing the phase mismatch. Even though a phase mismatch is present, it will not effect the measurement statistics since the measurements are done in the logical basis. We will obtain the result $ \ket{0} $ with a probability of $ |\alpha|^2$ and $\ket{1}$ with probability of $|\beta|^2$, regardless of the mismatch. Observing the measurement statistics remain undisturbed by the phase mismatch is highly beneficial for us. However, when a measurement isn't to be performed and the identity operation is  applied, the effect of the mismatch will inevitably be present. The qubit will continue on its path but the mismatch will negatively effect the algorithm.

At the current state of on-chip quantum computation, the proposed scheme will not be of substantial use, as the detectors are coupled to the chip externally. The advantages of this scheme become pronounced with the advent of integrated photodetectors on-chip \cite{sprengers2011waveguide}. The scheme will allow for measurements at the desired instant and position, which will decrease the decoherence, losses and cross-talk the qubit will be subjected to as it travels to the end of the circuit. The qubit will be directly measured at the integrated photodetector, which will be at a much closer proximity than the externally coupled photodetectors. Intermediate measurement results will be present in such a scheme, possibly enabling feedback before the finalization of the algorithm.

\section{Discussion}

Composed of highly general components, our gate may be implemented on various platforms on-chip. Demonstrations of extremely high fidelity RBS and MZI with extinction ratios exceeding 60 dB make silicon photonics a promising platform for the implementation of the proposed gate in a scalable manner \cite{Harris2017,Thompson2012}. Since MZI and phase shifters are fundamental components of on-chip quantum circuits, it is reasonable that we will observe progress in the fidelities and compactness of these devices. Furthermore, the presence of extremely low-loss, CMOS compatible, compact waveguide crossings with $9 \times 9$ $\mu\text{m}^2$ footprint and 0.03dB loss is a great advantage \cite{Ma2013}. Since waveguide crossings with improved performances are essential for photonic interconnects, significant advancements on this area can be expected \cite{wcinterconnect}.

Furthermore, there exists CMOS compatible, compact, low-loss, Bloch wave supporting, cascaded waveguide crossing arrays with losses around 0.02dB per crossing \cite{blochwc}. These structures may be used for the implementation of architecture specific multiport quantum circuits in a robust manner. For example, the implementation of the Quantum Fourier Transform requires the reversal of the order of the qubits to finalize the algorithm \cite{Nielsen2002}, this reordering may be universally  accomplished through the proposed SWAP gate. Nonetheless, utilizing the waveguide crossing array, the qubits may be directly swapped deterministically. If a reconfigurable operation is required, first and the final qubit basis may be brought next to one another and swapped. This process may be repeated for all the pairs via multiport waveguide crossing arrays. Such an application specific multiport architecture would decrease the number of RBS structures required for the reordering.

\section{Conclusion}

We have demonstrated the operation of a simple, reconfigurable and deterministic SWAP gate. We have discussed the physical issues related to the implementation of our gate and suggested possible ways to overcome emerging obstacles. Considering the possibility of integrated photodetectors, we have proposed a scheme which utilizes the proposed SWAP gate to perform selective measurements, presenting possible improvements regarding coherence and reconfigurability. We have examined previously produced silicon photonics structures which may be used in the physical implementation of the gate, possibly bringing forward alternative applications. We have  emphasized the crucial relationship between on-chip linear optical quantum computation and silicon photonics.

Significantly reducing the overhead associated with the SWAP gate, our proposal can be used to implement general algorithms on-chip for the immense goal of universal quantum computation. 

\bibliographystyle{apsrev4-1}

\end{document}